\documentclass[%
 preprint,
superscriptaddress,
 amsmath,amssymb,
 aps,
prb,
floatfix,
]{revtex4-1}

\usepackage{graphicx}
\usepackage{amssymb}

\usepackage{lineno}
\usepackage{subcaption}
\usepackage{color}

\usepackage{amsmath}
\usepackage{algorithm}
\usepackage[noend]{algpseudocode}
\usepackage{titlesec}

\titleformat*{\section}{\large\bfseries}
\titleformat*{\subsection}{\normalsize\bfseries}
\titleformat*{\subsubsection}{\normalsize\itshape}

\begin{document}



\title{Phase Segmentation in Atom-Probe Tomography Using Deep Learning-Based Edge Detection}

\author{Sandeep Madireddy}
\affiliation{Mathematics and Computer Science Division, Argonne National Laboratory, Lemont, IL 60439}

\author{Ding-Wen Chung}
\affiliation{Department of Materials Science and Engineering, Northwestern University, Evanston, IL 60208}

\author{Troy Loeffler}
\author{Subramanian K.R.S. Sankaranarayanan}
\affiliation{Nanoscience and Technology Division, Argonne National Laboratory, Lemont, IL 60439}

\author{David N. Seidman} 
\affiliation{Department of Materials Science and Engineering, Northwestern University, Evanston, IL 60208}
\affiliation{Northwestern University Center for Atom-Probe Tomography, Northwestern University, Evanston, IL 60208}

\author{Prasanna Balaprakash}
\affiliation{Mathematics and Computer Science Division, Argonne National Laboratory, Lemont, IL 60439}

\author{Olle Heinonen}

\affiliation{Materials Science Division, Argonne National Laboratory, Lemont, IL 60439}
\affiliation{Northwestern-Argonne Institute of Science and Engineering, 
Evanston, IL 60208}
\email[Corresponding author:]{heinonen@anl.gov}

\begin{abstract}

Atom-probe tomography (APT) facilitates nano- and atomic-scale
characterization and analysis of microstructural features. Specifically, APT is well
suited to study the interfacial properties of granular or heterophase systems. Traditionally,
the identification of the interface between, for precipitate and matrix phases, in APT
data has been obtained either by extracting iso-concentration
surfaces based on a user-supplied concentration value or by manually perturbing the concentration value until the iso-concentration surface qualitatively matches the interface. These approaches are 
subjective, not scalable, and may lead to inconsistencies due to local composition
inhomogeneities.

We propose a digital image segmentation approach based on deep neural networks
that transfer learned knowledge from natural images to automatically segment
the data obtained from APT into different phases. This
approach not only provides an efficient way to segment the data 
and extract
interfacial properties but does so without the need for expensive
interface labeling for training the segmentation model.

We consider here a system with a precipitate phase in a matrix and with three
different interface modalities---layered, isolated, and interconnected---that
are obtained for  different relative geometries of the precipitate phase. We
demonstrate the accuracy of our segmentation approach through qualitative
visualization of the interfaces, as well as through quantitative comparisons
with  proximity histograms obtained by using more traditional approaches.

\end{abstract}

\maketitle

\section{}
\label{Sec:Intro}


Advances in atom-probe tomography (APT) allow three-dimensional atomic
reconstruction of materials with an unparalleled spatial and atomic
resolution~\cite{Seidman:2007fu,Larson:2013ef,Coakley:2018cl}. Application of
APT to various materials, for example, metals, ceramics, biominerals and
composites, now provides atomic structure-property relations to facilitate
further materials development. APT is particularly useful in
studying interfacial properties of precipitates, surfaces and thin films.
Once the interface is identified, the elemental distribution along the
interface may be closely examined by using various statistical tools, such as
proximity histograms~\cite{Hellman:2000ca} and elemental
mapping~\cite{Felfer:2015ey}.

Traditionally, interfaces are identified through iso-concentration surfaces
constructed based on the marching cubes algorithm, which extracts an iso-concentration surface from a discrete scalar field with user-supplied
concentration values~\cite{Lorensen:1987kl}. The method is subjective and
prone to inconsistencies due to local composition inhomogeneities. In addition,
such a labor-intensive manual process prevents analyses of large amounts of APT datasets,
limiting the scope of APT studies.

In this paper, we focus on identifying the interface in a precipitate-matrix
system by phase segmentation.  This approach holds the potential to expedite and reduce
inconsistencies in the process of identifying interfaces and study of
interfacial properties and furthermore can be scaled up to high-performance
computer platforms.

Segmentation is an approach used to partition two- or three- dimensional
space into visually distinct and homogeneous regions with respect to certain
properties. Segmentation is widely studied in the context of digital images,
where the spatial information is represented by means of picture elements
(pixels) in two-dimension and volume elements (voxels) in three dimensions. A
pixel is the smallest unit of information that makes up a digital image in two-dimension; it is usually a square, and pixels are typically arranged in a two-dimensional grid. The color intensity in each pixel is represented by three
channels, representing the intensities of red, blue and green (commonly referred
to as RGB), respectively, and their combination uniquely defines the intensity
of the pixel. The attributes of the voxel are similar to that of the pixel
except that it is in three dimensions.

Classical approaches to segmentation include those based on intensity-thresholding based,
edge or boundary-detection based, region/similarity, clustering and 
graphs approaches (see~\cite{pal1993review} for a detailed review).
These approaches are primarily unsupervised; the segmentation models are
obtained from datasets consisting of image data without any explicitly labeled
segments. Several supervised segmentation approaches also exist that use {a
priori} knowledge involving the ground truth of the segments to recognize and
label the pixels in a new image according to one of the object classes on which the
model is trained. This approach is usually referred to as semantic
segmentation~\cite{garcia2017review} and tends to identify different
objects present in an image as well as their location.

The success of deep learning approaches in surpassing human-level accuracy
in tasks such as images classification~\cite{he2015delving} and language
translation~\cite{wu2016google} has motivated several recent works in the
field of digital image segmentation. This research typically has
focused on the semantic segmentation approach, since it has the potential to achieve
complete image/scene understanding, which is a crucial aspect of computer
vision. There are several applications under the umbrella of computer vision,
such as autonomous transport~\cite{teichmann2016multinet} and human-computer
interaction, as well as other applications, such as medical image
analysis~\cite{litjens2017survey} and remote
sensing~\cite{kampffmeyer2016semantic}, that have adopted semantic
segmentation. One shortcoming of this approach is that it is applicable
for segmenting only the objects (with a distinct shape) used in training sets.

Alternatively, edge (contour) detection has been significantly improved
with deep learning approaches~\cite{zhang2017recent}. A supervised learning
approach is used for edge detection, wherein each pixel is labeled as either edge
or nonedge. This approach is slightly different from semantic segmentation in
that there are only two classes (edge and noedge) and less semantic
knowledge; hence this approach could be applicable for segmentation of objects
with morphologies different from those in the training data. In the case of APT, 
the precipitate shape can range from a thin
slab to a complicated irregular volume; it is therefore a compelling case for the application of
edge detection to segment the precipitate from the matrix.

Early deep learning approaches for edge detection used a conventional
convolutional neural network (CNN)~\cite{hwang2015pixel}. Later
approaches replaced the CNN with fully convolutional networks (FCNs), which
provide an end-to-end framework for pixelwise label
prediction~\cite{long2015fully}. The holistically nested edge detection (HED)
approach~\cite{xie2015holistically} was subsequently proposed, which utilizes FCN along
with the side outputs (model predictions at the intermediate layers of the network) 
and deep supervision to significantly improve the edge
detection. HED was also demonstrated to achieve human-level accuracy in edge detection. Several enhancements
have been proposed for HED~\cite{kokkinos2015pushing, liu2017richer}, but it 
remains the most widely used approach because of its efficiency and multiscaling 
scheme to handle resolution and scale problems~\cite{zhang2017recent};
hence, we adopt HED for segmentation in this work. 

Although deep learning approaches have been shown to be successful in many
tasks, one shortcoming is that they rely heavily on being provided with
precise and abundant data to train the deep neural networks underlying the
approaches. Especially for the supervised learning used for edge detection,
large amounts of labeled data are required. Although labeled data are abundantly
available for edge detection in natural images, collecting such data is challenging
in problems such as interface detection in APT because significant
time and effort are required to conduct each experiment \cite{Larson:2013ef} and to
manually identify the iso-concentration surfaces as labeled data. We circumvent
this challenge of labeled data collection for training the edge detection
model by adopting transfer learning, which in general seeks
to generalize a model trained on one task to another similar task. 
More specifically, the transfer learning approach we adopt utilizes the knowledge
acquired from learning edge detection features on the source domain (natural
images), which has abundant labeled data, for a target domain (APT) edge
detection. This transfer learning approach can also be readily generalized to other imaging techniques, such as analysis of x-ray or transmission electron microscopy tomography, as well as to the analysis of synthetic data obtained from, e.g., molecular dynamics simulations, where edge detection of features such as grain boundaries is a central component.

In our work, we present a digital image segmentation based surface extraction as an alternative to manual and,
{ad hoc} construction of iso-concentration surfaces for APT datasets. We show that
our approach can transfer learn edge detection features from natural
images to segment APT reconstructions accurately and efficiently. We demonstrate the
qualitative and quantitative accuracy of our approach using a synthetic dataset 
constructed with molecular dynamics (MD) simulations, as well as experimental APT datasets
of Co and Al superalloys \cite{Bocchini:2018}.

\section{Results}
\label{Sec:methods}

\subsection{Holistically-Nested Edge Detection}

We adopt HED, an end-to-end edge
detection approach that performs image-to-image prediction (i.e., takes an image
as input, and outputs the prediction at each pixel) by means of a deep learning
model that leverages FCN and deeply
supervised nets. FCN is similar to the regular CNN model used for classification, 
but the last fully connected layer is replaced by another convolution layer 
with a large filter size, which allows pixelwise label prediction. 
Deep supervision is achieved by using the local output from each of the hidden
layers (analogous to the final output obtained from
a network truncated at the current hidden layer) and
back-propagating the error not only from the final layer but simultaneously
from all the local outputs in the learning stage. The side outputs and deep supervision contribute
to a significant performance gain over the patch-based CNN and simple FCN for
edge detection.

The training phase of this approach aims to learn a functional mapping
between the two-dimensional input image described by $N$ pixels $X_k$, $k=1,...,N$, 
where edge detection is desired, and the corresponding ground truth binary 
edge map ($Y_k$) on all the pixels of image $X_k$. This map is obtained by training a
neural network, which is composed of a VGGNet (a neural network consisting of 16
convolutional layers, five pooling layers, and three fully-connected layers, which was
proposed by the Visual Geometry Group (VGG)~\cite{simonyan2014very}), where the
fifth pooling layer and the fully connected layers are trimmed, resulting in five
stages with a total of 16 convolutional layers. The side output layer
is connected to the last convolutional layer in each stage for deep supervision.

The process of collecting APT data is expensive and time-consuming, and manually identifying the 
interfaces is cumbersome, labor-intensive and subjective. Hence, we adopt the transfer learning 
approach in which (1) the parameters for the trimmed VGGNet part of the network 
are initialized to the weights from VGGNet pretrained on the Imagenet
dataset~\cite{deng2009imagenet} and (2) the entire network is then trained by using
the Berkeley Segmentation Dataset and Benchmark (BSDS
500)~\cite{arbelaez2011contour} dataset (composed of 200 training, 100
validation, and 200 testing images) containing a wide variety of natural
scenes with at least one discernable object ({e.g.} birds, animals).
Each image in the dataset has been manually annotated to obtain the ground truth
contours. This training approach is similar to  the procedure for edge
prediction in natural images outlined by Xie and Tu
(2015)~\cite{xie2015holistically}. 

\subsection{Orthogonal Volumetric Segmentation}

The neural network trained by using the procedure mentioned in the preceding
section is then used to perform segmentation on a three-dimensional APT
dataset. 

Before segmentation, the data have to be prepared and processed into a
suitable format. The datasets obtained from APT consist of the spatial coordinates
of each atom and a label for their chemical identity. These data are
transformed into a regular 3D grid of atomic concentrations, where the grid is
obtained by partitioning the 3D space into a series of voxels, and the
concentration is calculated for a given chemical species on each voxel based
on its relative atomic fraction with respect to the other species. This grid along
with the concentrations is hereafter referred to as the \texttt{concentration space}. 
The concentration values range between $[0,1]$; hence they can be converted 
into grayscale images with a single intensity channel, and subsequently replicate
these intensity values into three channels corresponding to RGB of 
each voxel.
The segmentation approach can then be applied to the \texttt{concentration space}
and its associated RGB voxels.

We propose to segment the \texttt{concentration space} by extracting 2D
slices in each of the three orthogonal directions and detecting the
edges using the HED model trained on natural images. Once the edges
are obtained on each of the image slices in the three orthogonal directions, a
3D edge detection map for each slice direction is obtained by stacking all the 2D edges
in that direction. Then all the three 3D edge detection maps are merged into one by
taking a union of all the voxels detected to have edges. Since the APT datasets studied here have only
two phases, the obtained edge detection map in 3D serves
as the interfacial surface between the two phases. We note, however that the
thickness of this surface depends on the thickness of the edges delineating
the two surfaces. If the edges are thicker and
extend to more than one voxel, there could be two surfaces, one on either side
of the edge. This configuration is still acceptable, however since it shifts the proximity
histograms (described in the Methods section) by only a small distance from the
interface and hence does not have significant effects on the interface properties
calculations.

\subsection{Segmentation}

We evaluate the effectiveness and quantitative fidelity of the proposed
supervised edge detection and transfer-learning-based digital image
segmentation approach using three interface modalities: (1) layered interface,
in which the precipitate and matrix are two different layers separated by a
thin interface; (2) isolated interface, in which the precipitate is an
ellipsoid embedded in the matrix; and (3) interconnected interface, which is a
general case where the precipitate phase  exhibits an irregular morphology.
The layered interface structures examined are synthetic and were generated by
using molecular dynamics simulations or using IVAS software. The isolated and
interconnected precipitates APT datasets were obtained experimentally.

\subsubsection{Layered Structures}

\begin{figure}[h!]
    \centering

    \begin{subfigure}[t]{0.31\textwidth}
        \centering
        \includegraphics[height=\linewidth]{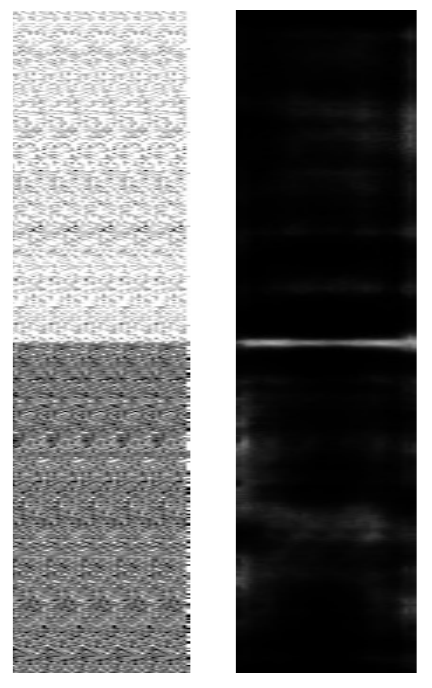}
        \caption{\texttt{Concentration space} (left) and edge detection map (right) of a slice along 
        the Y-direction.}
        \label{fig:layer_yslice}
    \end{subfigure}
    ~ 
    \begin{subfigure}[t]{0.31\textwidth}
        \centering
        \includegraphics[height=\linewidth]{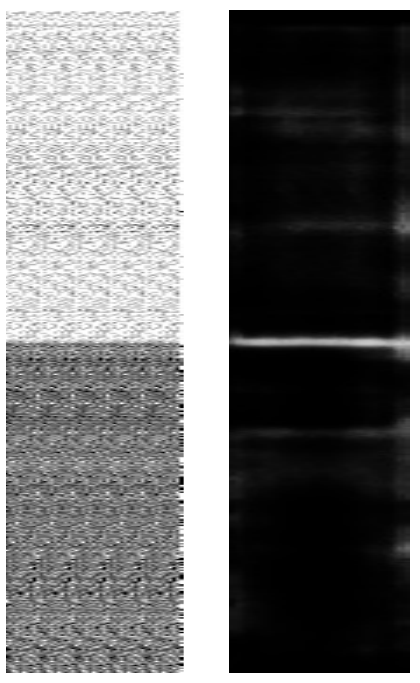}
        \caption{\texttt{Concentration space} (left) and edge detection map (right) of a slice along 
        the Z-direction.}
        \label{fig:layer_zslice}
    \end{subfigure}  
    ~
    \begin{subfigure}[t]{0.31\textwidth}
        \centering
        \includegraphics[height=\linewidth]{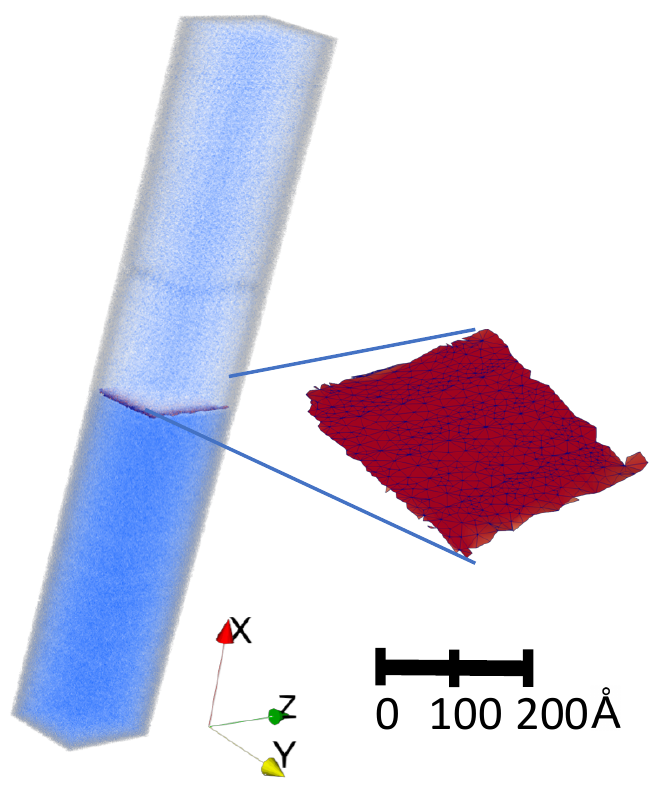}
        \caption{Edge surface obtained by merging the edges in the Y- and Z-directions.}
        \label{fig:layer_full3d}
    \end{subfigure}%
    \caption{Two- and three-dimensional edge detection map in a synthetic Co-Al alloy generated from MD simulation.}
\end{figure}

A synthetic layer structure was generated by using the MD
approach outlined in Section~\ref{Sec:Data_prep_MD}. The \texttt{Concentration space} was obtained for the Co atoms with a
voxel size chosen as $2\times2\times2$~\AA$^3$. In this particular case, there was no
interface component in the X-direction, so only the slices in the Y- and Z-
directions were used. A 2D slice in the Y-direction and the corresponding image
showing the edge detection map (indicated in white) is shown in
Figure~\ref{fig:layer_yslice}. The edge detection map correctly identified the interface
between the top and bottom regions. A similar exercise on a slice in the Z-direction is shown in Figure~\ref{fig:layer_zslice}, where the interface was
also captured appropriately by the edge detection map. Merging the 2D edge detection map
from slices in the Y- and Z direction, produces the surface separating the top and bottom
regions as shown in Figure~\ref{fig:layer_full3d}.

\begin{figure}[h!]
    \centering

    \begin{subfigure}[t]{0.45\textwidth}
        \centering
        \includegraphics[height=0.44\linewidth]{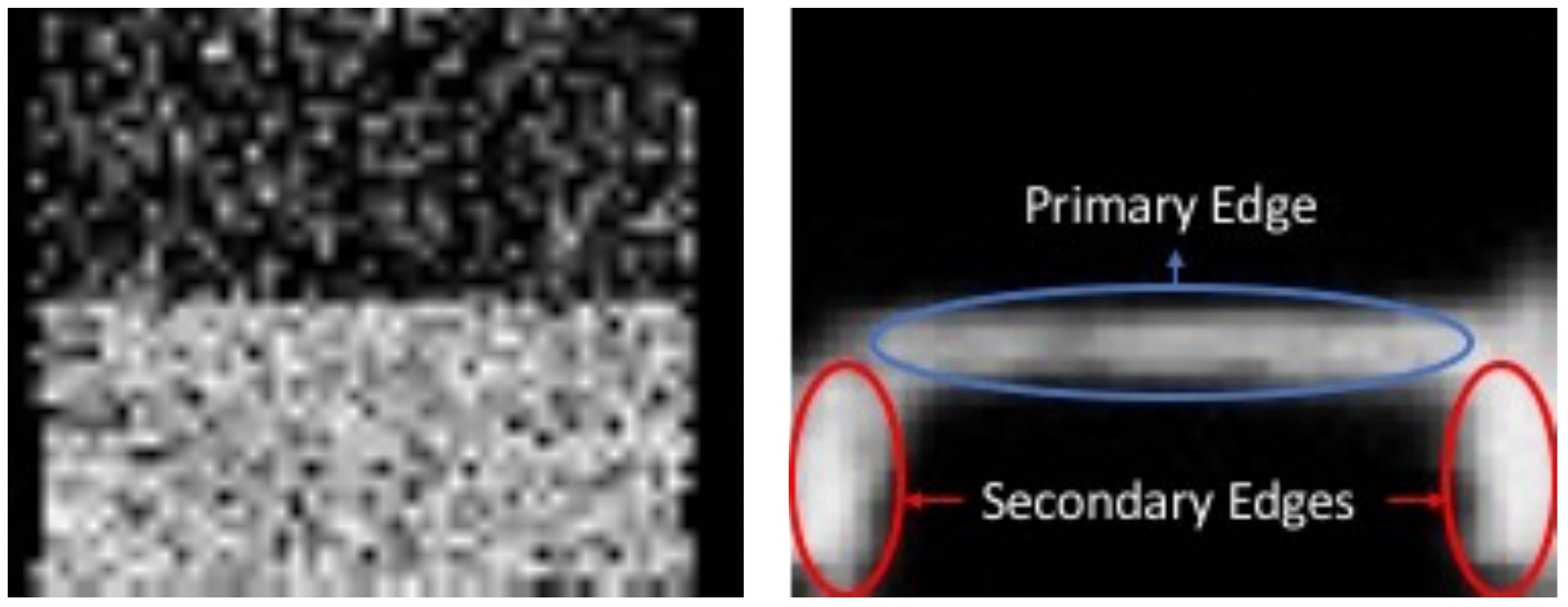}
        \caption{\texttt{Concentration space} (left) and edge detection map (right) of a slice along 
        the Y-direction. The desired primary edge separating the two layers as well as the secondary
        edges is highlighted.}
        \label{fig:layer2_yslice}
    \end{subfigure}
    ~ 
    \hfill
    \begin{subfigure}[t]{0.45\textwidth}
        \centering
        \includegraphics[height=0.44\linewidth]{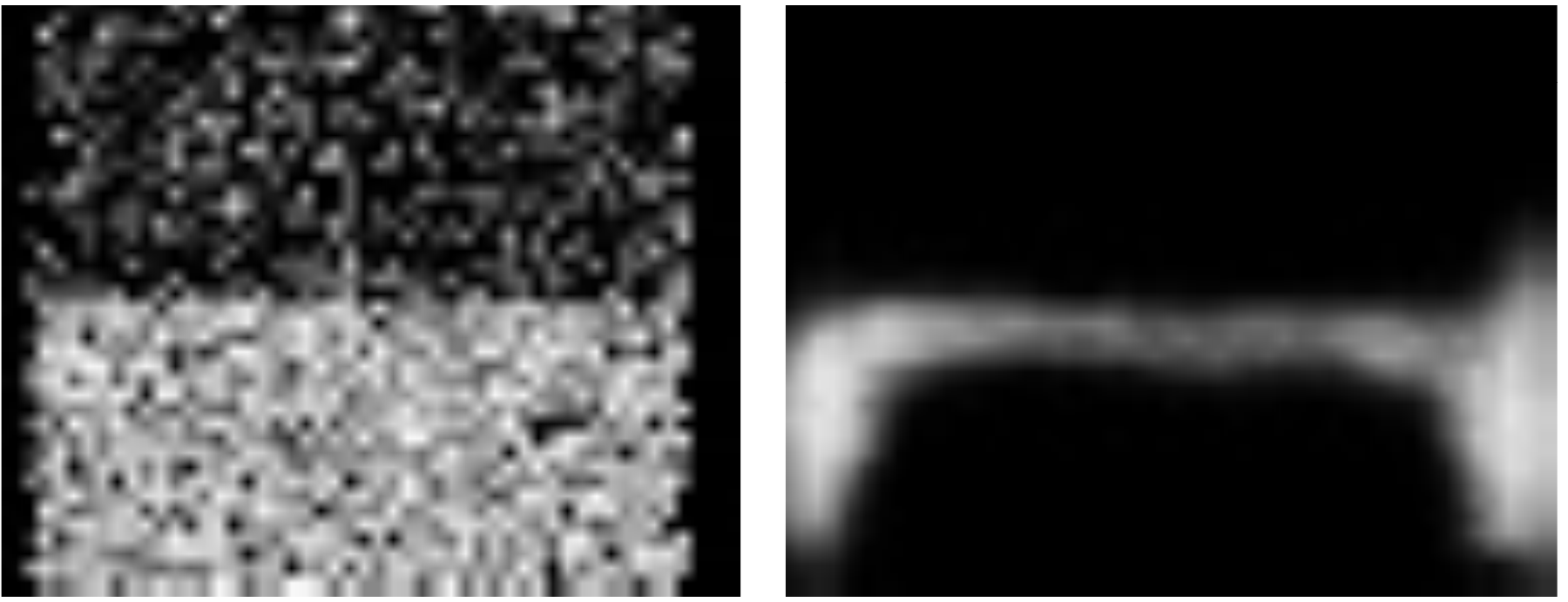}
        \caption{\texttt{Concentration space} (left) and edge detection map (right) of a slice along 
        the Z-direction.}
        \label{fig:layer2_zslice}
    \end{subfigure}  
    ~
    \begin{subfigure}[t]{\textwidth}
        \begin{minipage}{0.5\linewidth}
        \centering
        \includegraphics[height=0.7\linewidth]{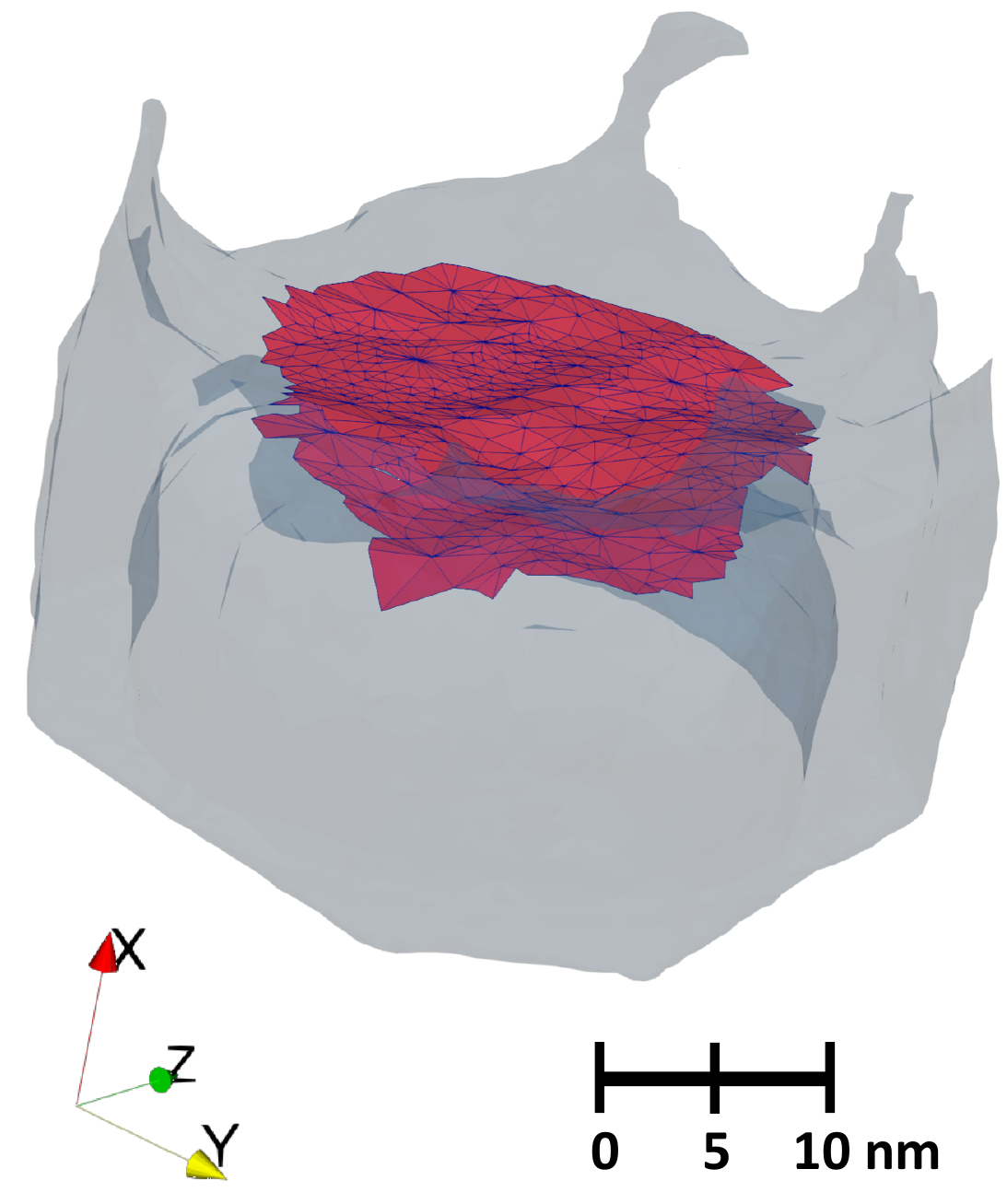}
        \end{minipage}
        \begin{minipage}{0.45\linewidth}
        \caption{Edge surfaces extracted after merging the edges in Y- and Z directions.
        }
        \end{minipage}
        \label{fig:layer2_full3d}
    \end{subfigure}%
    \caption{Two- and three-dimensional edge detection map in synthetic Co-Al-W alloy generated by using IVAS.}
    \label{fig:IVAS_syntheticCoAlW}
\end{figure}

Similarly, a synthetic structure of a Co-based superalloy generated by using IVAS is shown in Figure~\ref{fig:IVAS_syntheticCoAlW}. The synthetic structure contains a layer of $\gamma^\prime$(L1$_2$)
precipitate and a layer of $\gamma$(fcc) matrix.
This dataset contains additional spatial variations
compared with the MD structure. As in the previous
case, 2D slices were generated only in the Y- and Z directions to obtain the edge
detected surfaces. A slice in the Y-direction and the corresponding edge detection map
are shown in Figure~\ref{fig:layer2_yslice}, and corresponding images for a slice
in the Z-direction are displayed in Figure~\ref{fig:layer2_zslice}.  The total
number of atoms from all the elements Co, Al and W was 1,983,127, with the dimensions of the enclosing volume being $41\times
51\times51$~nm$^3$. The \texttt{Concentration space} was obtained for the Co atoms
with a voxel size chosen as $1\times1\times1$~nm$^3$. We note that unlike the
previous case, the detected edges are thicker, and we observe secondary edges
on the side (Figure~\ref{fig:layer2_yslice}). The thickness of the edges is a consequence of the uncertainty in
the boundary between the top and bottom layers as well as the HED edge detection
algorithm itself, which tends to produce thicker edges~\cite{wang2017deep}.
Other edge detection approaches such as the crisp edge detection
(CED)~\cite{wang2017deep} could be used to alleviate this situation and
decrease the thickness of the edges. The consequence of a thicker edge is only
that we would obtain two surfaces, one at the top and the other at the bottom as
displayed in Figure~\ref{fig:IVAS_syntheticCoAlW}. This does not affect the
accuracy of the proximity histogram, or \textit{proxigram} (see Methods section), however, because it shifts the \textit{proxigram} only by a
small distance from the interface (see Section~\ref{res:proxigram} for a discussion on
the \textit{proxigrams} obtained). The secondary edges detected on the side  
could be removed by preprocessing the 2D slices to remove the empty volume.

\subsubsection{Isolated Phase}

To further validate the effectiveness of the implemented segmentation method, we used
an experimental APT dataset of a L1$_2$ strengthened Al-Er-Sc-Zr-Si superalloy. The interface exhibits a more complex morphology with the precipitate having
an ellipsoidal morphology.

\begin{figure}[t!]
    \centering

    \begin{subfigure}[t]{0.475\textwidth}
        \centering
        \includegraphics[width=\linewidth]{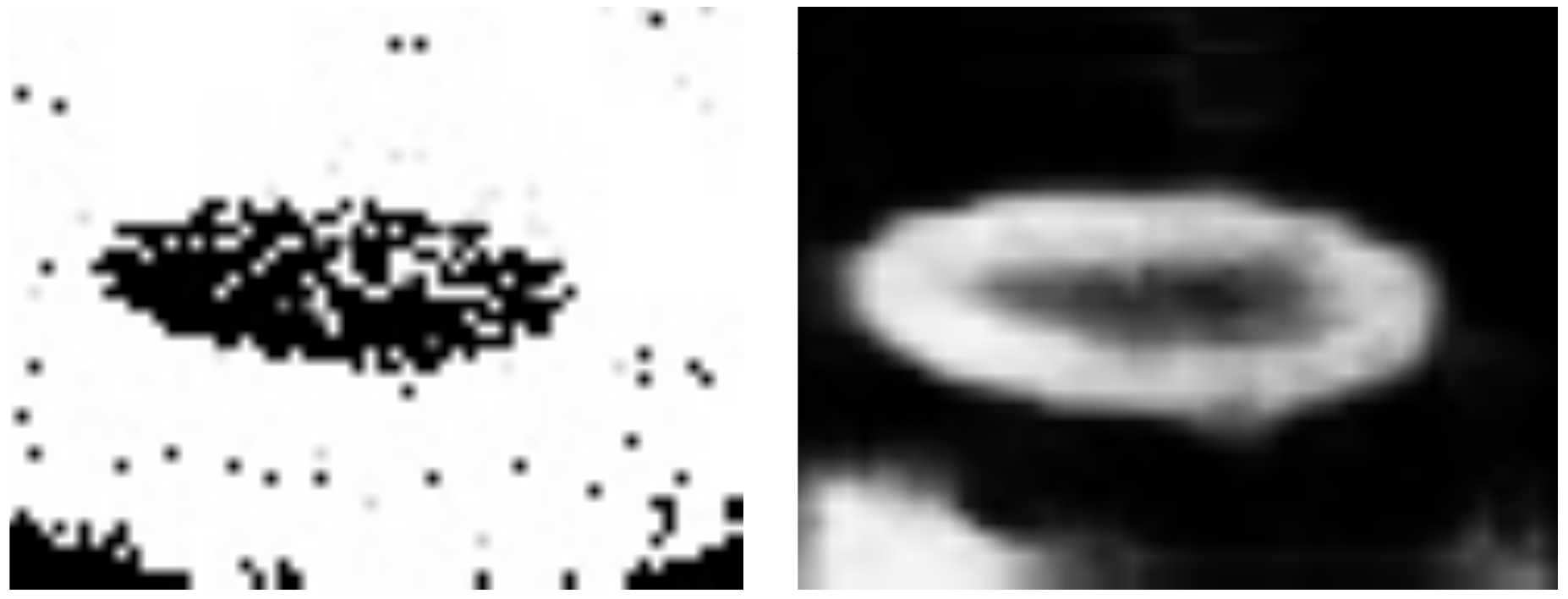}
        \caption{\texttt{Concentration space} (left) and edge detection map (right) of a slice along 
        the X-direction.}
        \label{fig:inclusion_xslice}
    \end{subfigure}
    ~ 
    \begin{subfigure}[t]{0.475\textwidth}
        \centering
        \includegraphics[width=\linewidth]{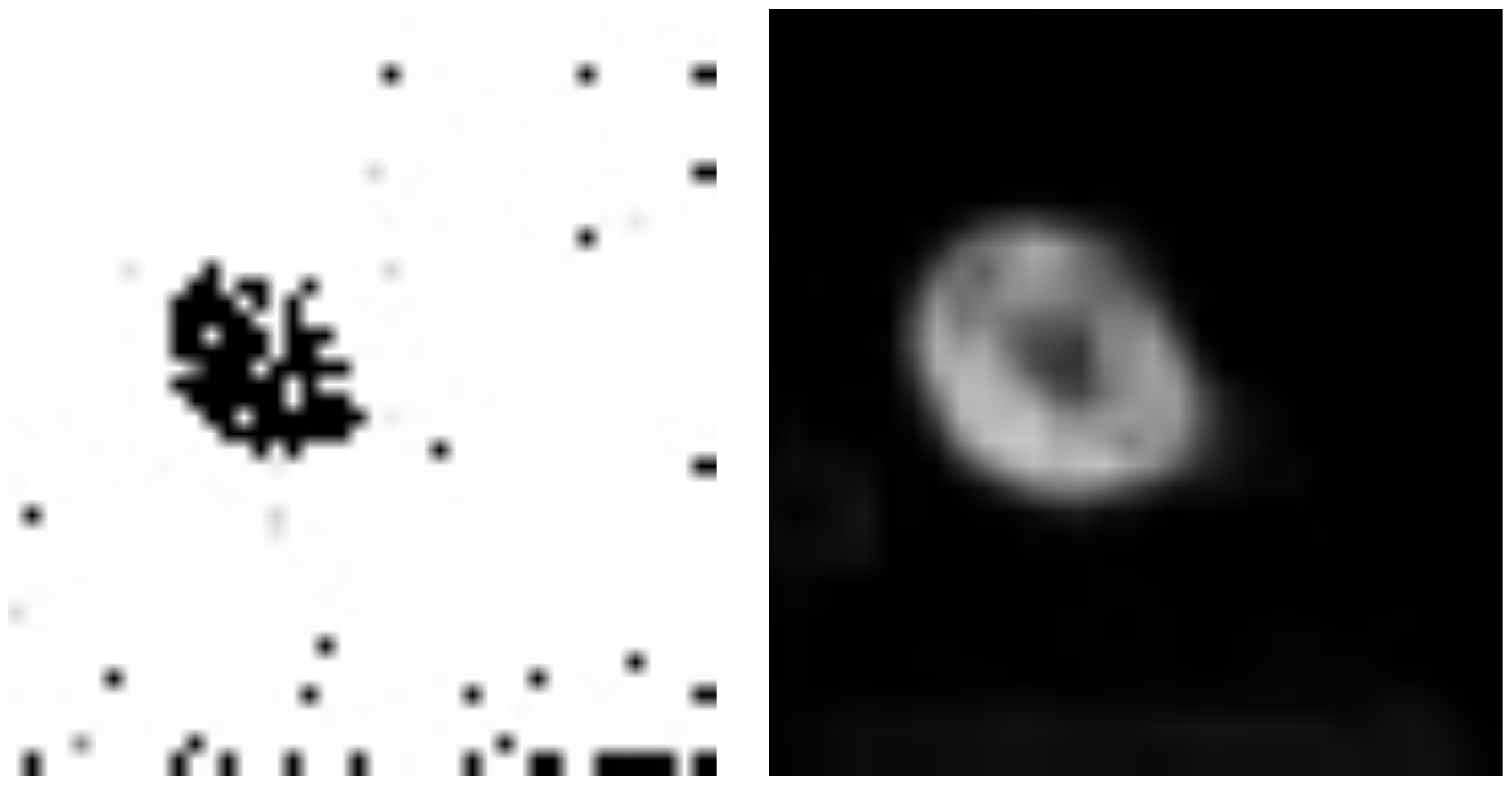}
        \caption{\texttt{Concentration space} (left) and edge detection map (right) of a slice along 
        the Y-direction.}
        \label{fig:inclusion_yslice}
    \end{subfigure}    
    ~ 
    \begin{subfigure}[t]{0.475\textwidth}
        \centering
        \includegraphics[width=\linewidth]{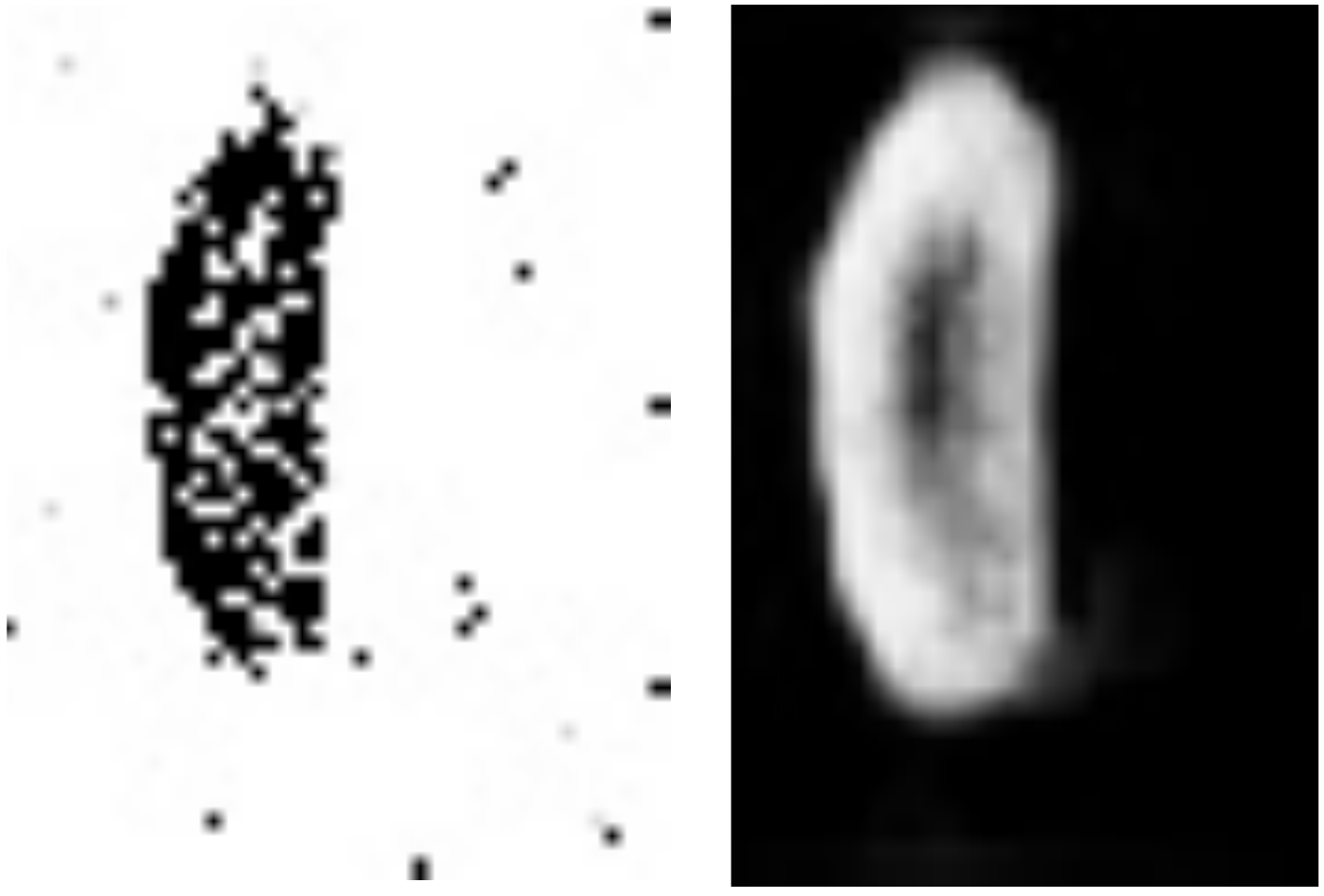}
        \caption{\texttt{Concentration space} (left) and edge detection map (right) of a slice along 
        the Z-direction.}
        \label{fig:inclusion_zslice}
    \end{subfigure}    
    ~
    \begin{subfigure}[t]{0.475\textwidth}
        \centering
        \includegraphics[width=\linewidth]{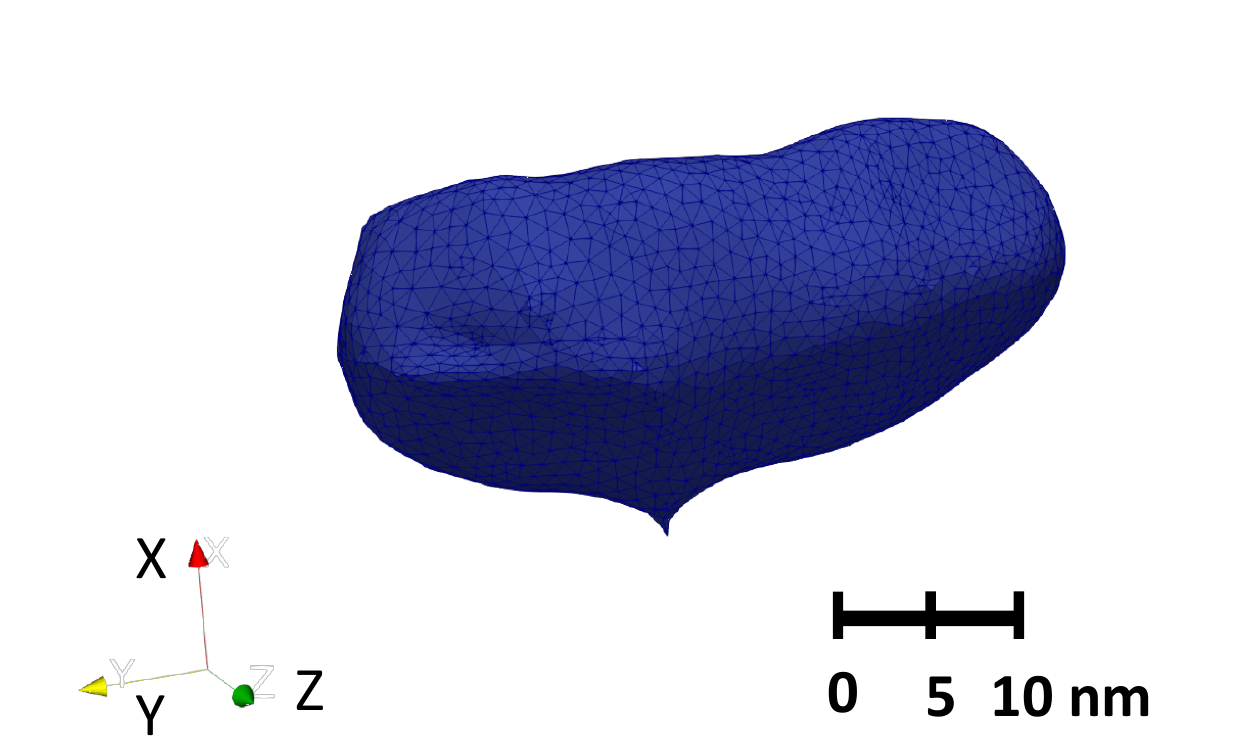}
        \caption{The edge surface obtained by merging the edges on slices in the X-,Y-, and Z-direction.}
        \label{fig:inclusion_full3d}
    \end{subfigure}%
    \caption{Two- and three-dimensional edge detection map in the experimentally obtained APT data set from an L1$_2$-strengthened Al-Si-Sc-Er superalloy.}
\end{figure}

For the APT tomogram with a three-dimensional varying morphology, the 2D slices have to be extracted
in each of the three orthogonal directions to reconstruct accurately the
precipitate geometry. The 2D slices and the corresponding edge detection map in the X-, Y-, Z-directions are shown in Figures~\ref{fig:inclusion_xslice},
\ref{fig:inclusion_yslice}, and \ref{fig:inclusion_zslice}, respectively. The morphology
of the precipitate has been accurately retrieved as seen in the edge detection map
obtained on a 2D slice in each of the three orthogonal directions,
and the fully reconstructed 3D surface are obtained from fusing all slices in three directions
(Figure~\ref{fig:inclusion_full3d}). We observe the thick edges in the experimental APT dataset
as well. The ability to capture accurately the interface in the Al superalloy,
which contains tens of percents of alloying elements (Si, Sc, Zr, and Er), signifies
the interface-identifying capability of the HED method.

\subsubsection{Interconnected Phases}

\begin{figure}[t!]
    \centering

    \begin{subfigure}[t]{0.475\textwidth}
        \centering
        \includegraphics[width=\linewidth]{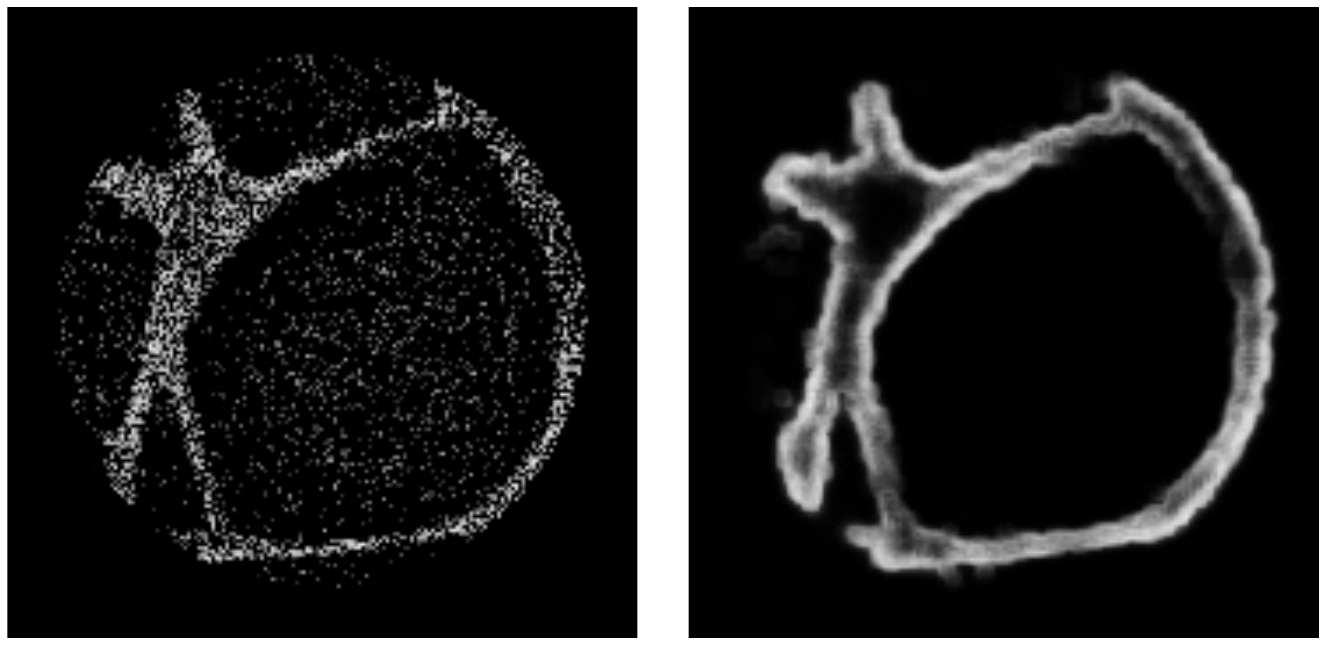}
        \caption{\texttt{Concentration space} (left) and edge detection map (right) of a slice along 
        the X-direction.}
        \label{fig:connected_xslice}
    \end{subfigure}
    ~ 
    \begin{subfigure}[t]{0.475\textwidth}
        \centering
        \includegraphics[width=\linewidth]{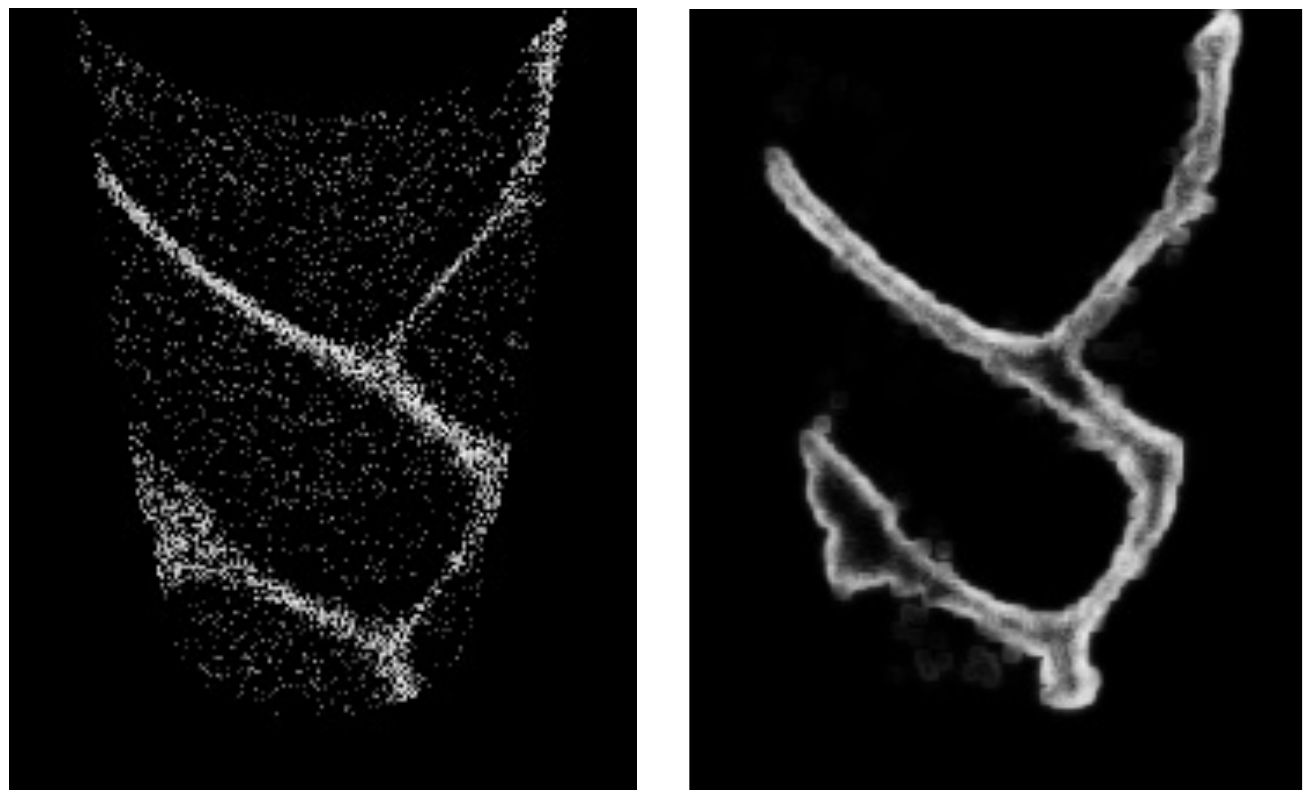}
        \caption{\texttt{Concentration space} (left) and edge detection map (right) of a slice along 
        the Y-direction.}
        \label{fig:connected_yslice}
    \end{subfigure}
    ~ 
    \begin{subfigure}[t]{0.475\textwidth}
        \centering
        \includegraphics[width=\linewidth]{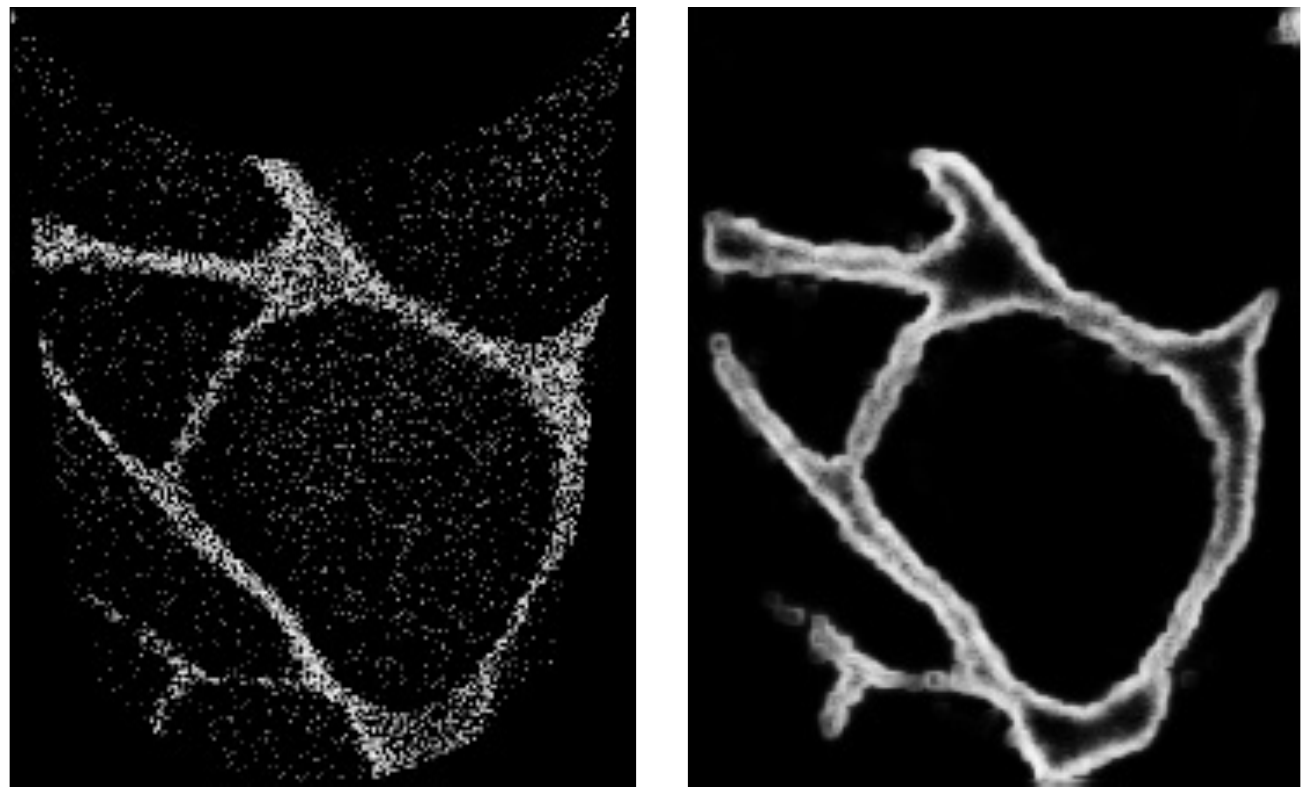}
        \caption{\texttt{Concentration space} (left) and edge detection map (right) of a slice along 
        the Z-direction.}
        \label{fig:connected_zslice}
    \end{subfigure}   
    ~ 
    \begin{subfigure}[t]{0.475\textwidth}
        \centering
        \includegraphics[width=0.49\linewidth]{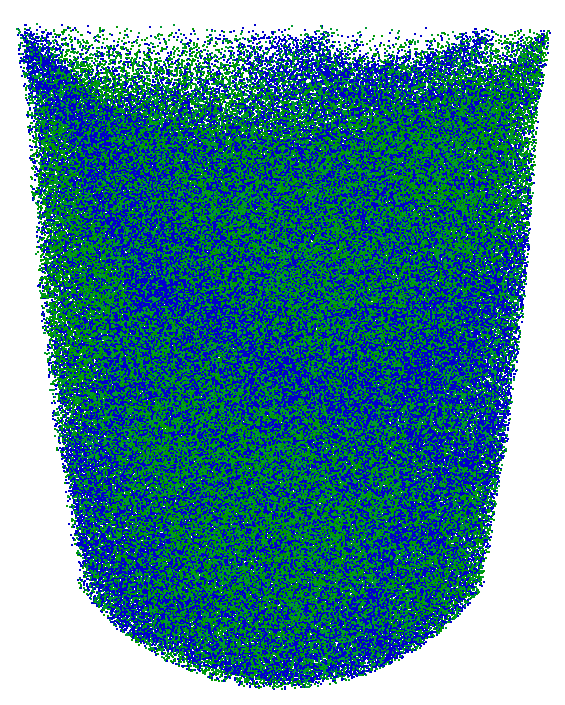}
                \includegraphics[width=0.49\linewidth]{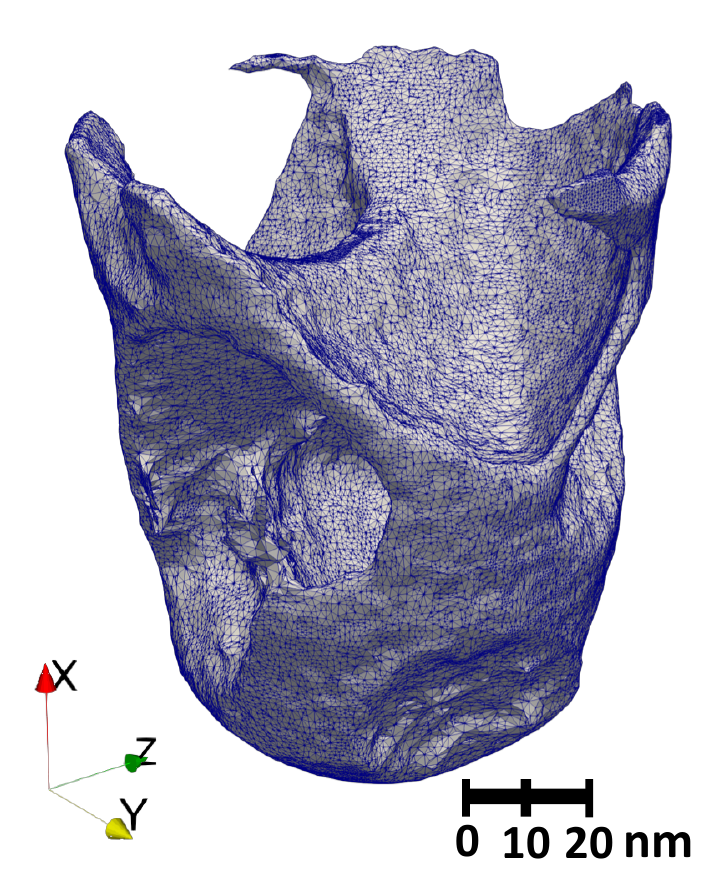}
        \caption{Three-dimensional point cloud reconstruction of the entire Co superalloy APT specimen (left) 
        and the resulting interface from the edge detection method (right). This method identifies 
        the interconnected $\gamma$ matrix channels on the order of tens of nm.}
        \label{fig:connected_full3d}
    \end{subfigure}%
    \caption{Two- and three-dimensional edge detection map of the experimentally obtained APT data set from a Co-based superalloy.}
    \label{fig:interconnectphase}
\end{figure}

Figure~\ref{fig:interconnectphase} shows the three-dimensional reconstruction
of the entire APT nanotip from a Co-based superalloy \cite{Bocchini:2018}.  
The narrow $\gamma$ matrix channel
and the interconnectivity of the $\gamma^\prime$ from precipitate coalescence
result in an overall complex interfacial structure. 

This is the most general case in which the precipitate is irregular and distributed throughout the examining volume.   
A 2D slice of the \texttt{concentration space} in the X-direction and
the corresponding edges detected are shown in Figure~\ref{fig:connected_xslice},
where the edge detection approach accurately identifies
and segments the $\gamma$ matrix phase from the $\gamma^\prime$ precipitate
phase. A 2D slice in the Y and Z-directions and their corresponding edge detection map are shown
in Figures~\ref{fig:connected_yslice} and  \ref{fig:connected_zslice},
respectively. Figure~\ref{fig:connected_full3d} shows the full 3D representation of the surface delineating the $\gamma$ and $\gamma'$ phases
that is obtained by fusing the information from the 2D slices in the three orthogonal directions.

\begin{figure}[t!]
    \centering

    \begin{subfigure}[t]{0.475\textwidth}
        \centering
        \includegraphics[width=0.75\linewidth]{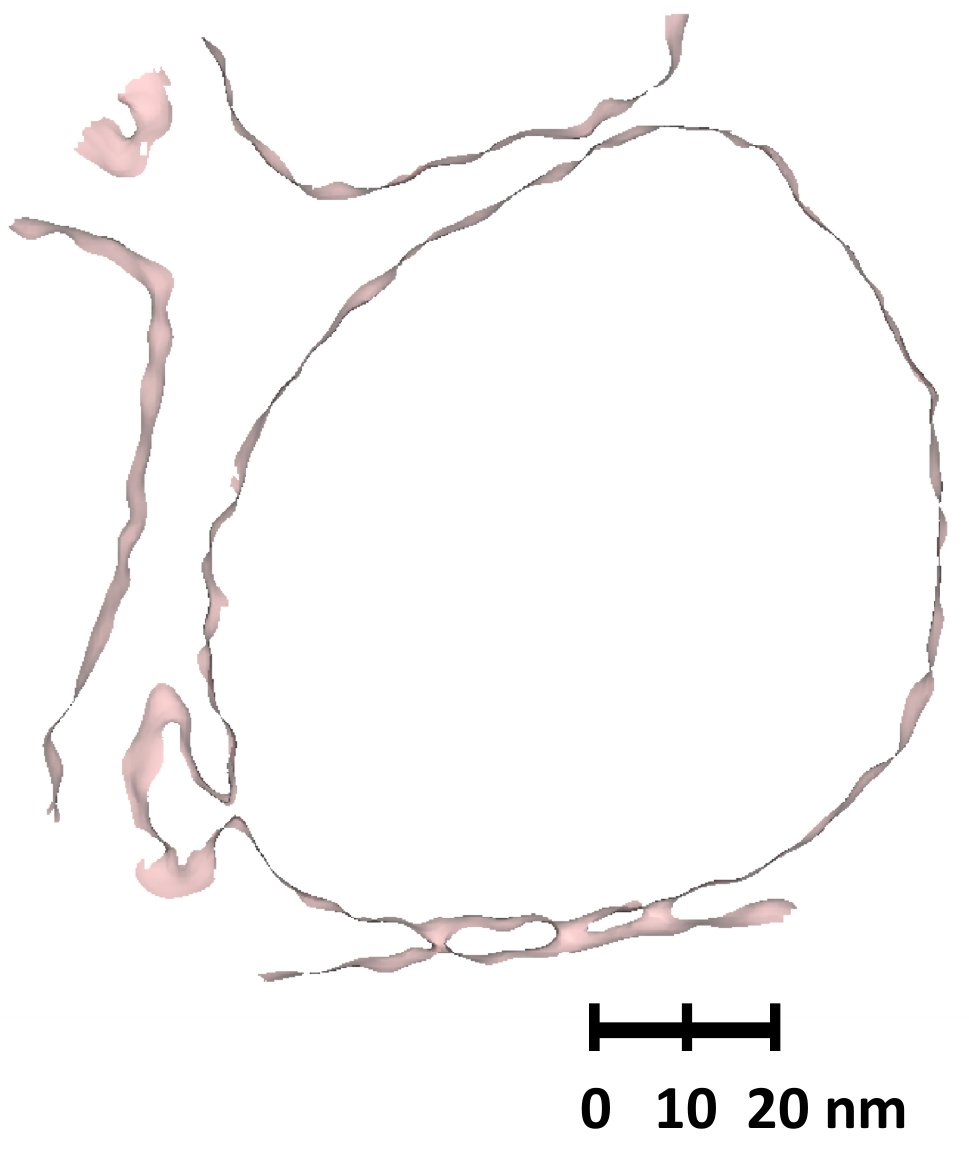}      

        \caption{Slice of the iso-concentration surface (Co = 0.855) along the X-axis obtained by using IVAS.}
        \label{fig:isoConc}
    \end{subfigure}
    ~
    \begin{subfigure}[t]{0.475\textwidth}
        \centering
        \includegraphics[width=\linewidth]{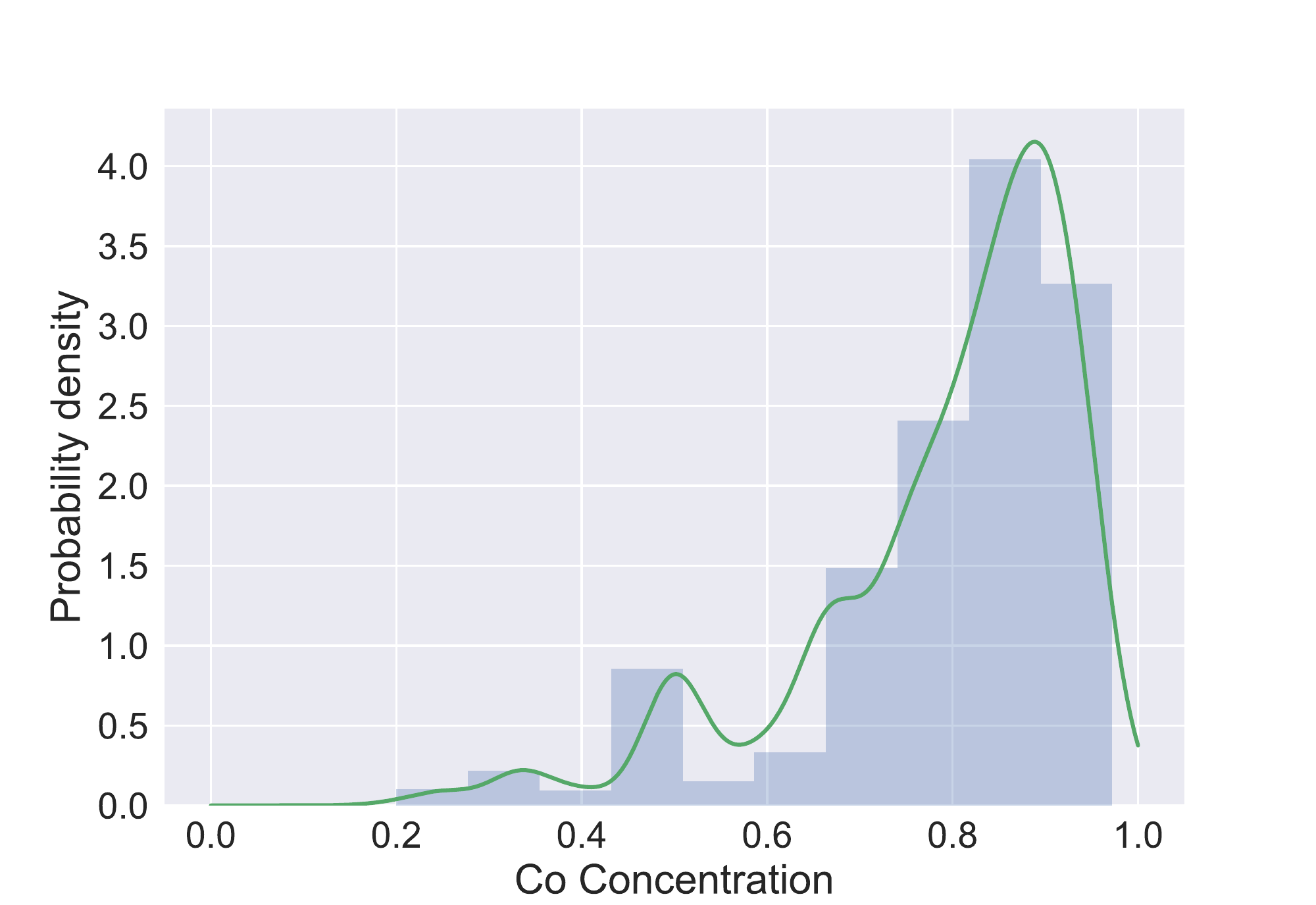}        
        \caption{Histogram of the concentrations at the pixels corresponding to the 
        edge detected map shown in Figure~\ref{fig:connected_xslice} (peak at Co = 0.87).}
        \label{fig:X_slice_hist}
    \end{subfigure}
    \caption{Comparison of iso-concentration interface generated by using IVAS software and the HED-generated interfaces.}
\end{figure}

A qualitative comparison of the \texttt{concentration space}
and the edge detection maps on each slice shows that the location of the matrix/precipitate interface is
fairly well captured by the edges. To make a quantitative comparison,
we obtained the iso-concentration surface using IVAS, by carefully tuning the Co concentration until the precipitate and matrix interface is captured. The obtained iso-concentration surface corresponds to a Co concentration of 0.855, and the detected surface corresponding to this concentration is shown in Figure~\ref{fig:isoConc}. With the HED approach, for the identical 2D slice, a histogram of the Co concentration at the HED detected edge is shown in Figure~\ref{fig:X_slice_hist} and has a peak at Co = 0.87, which agrees well with the iso-concentration surface obtained via IVAS, thus validating the interface detected by using the HED method against the ubiquitous iso-concentration surface approach.

\subsection{{Proxigram} Estimation}
\label{res:proxigram}

\begin{figure}[htp!]
    \centering

    \begin{subfigure}[t]{\textwidth}
        \centering
        \includegraphics[width=0.8\linewidth]{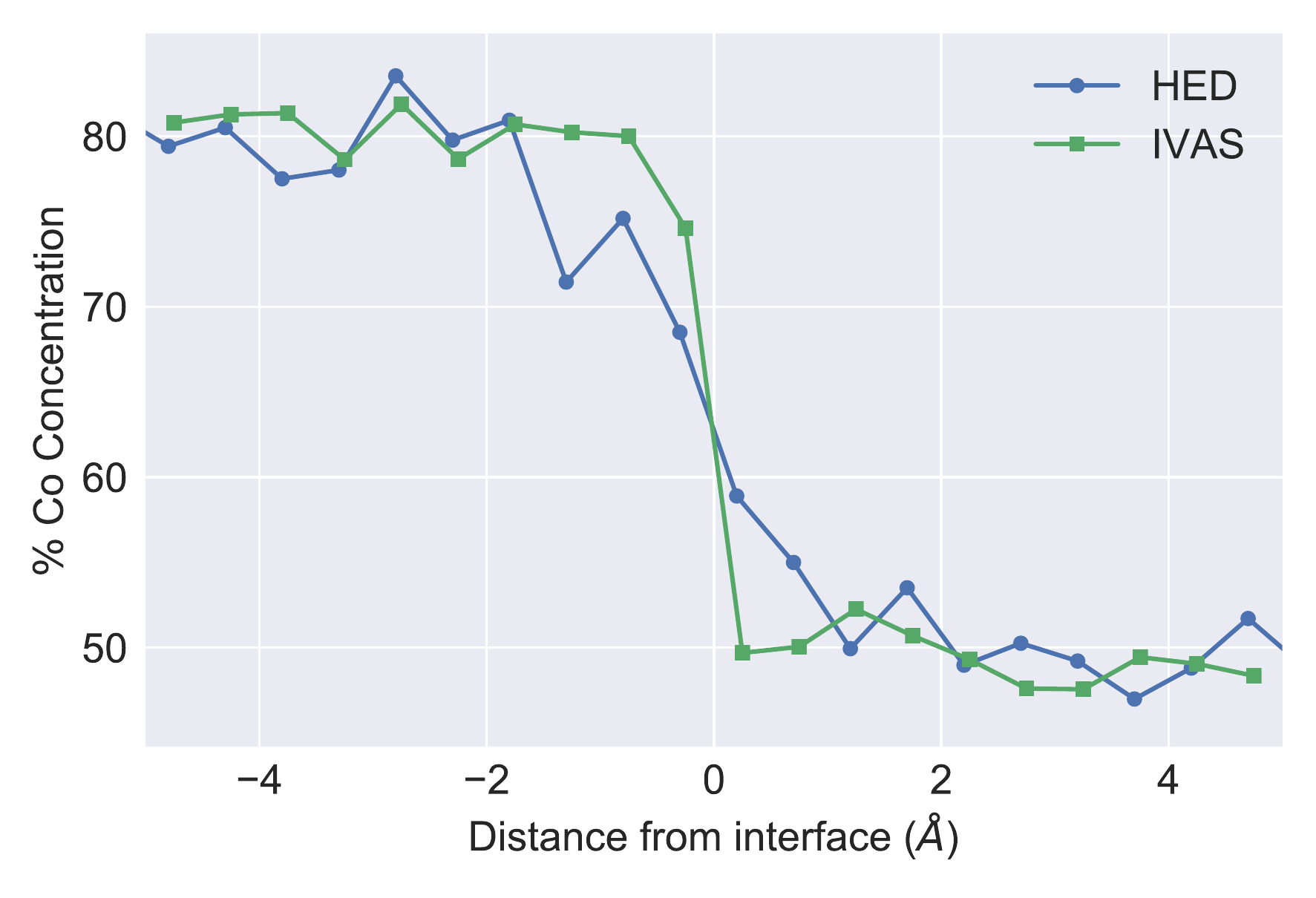}
        \caption{\textit{Proxigram} concentration profile from the synthetic MD layered structure using HED and IVAS.}
        \label{fig:MD_layered}
    \end{subfigure}
    ~ 
    \begin{subfigure}[t]{\textwidth}
        \centering
        \includegraphics[width=0.85\linewidth]{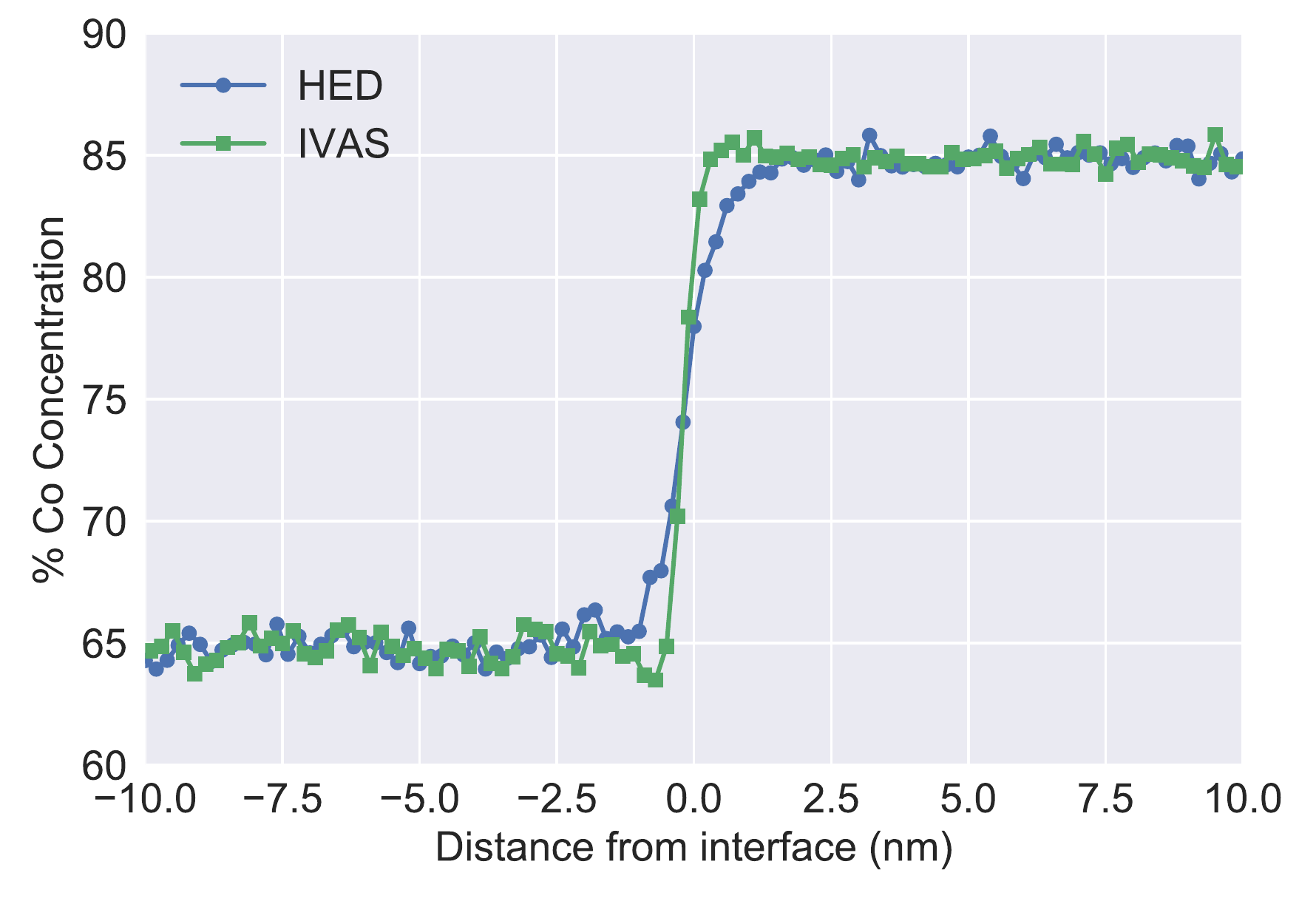}
        \caption{\textit{Proxigram} concentration profile obtained from the synthetic IVAS dataset using HED and IVAS.}
        \label{fig:3DAPT_layered}
    \end{subfigure}

    \caption{\textit{Proxigram} comparisons on the MD and IVAS generated synthetic datasets. The \textit{Proxigram} concentration profile derived from IVAS display a narrower interfacial compared with the HED method.}
\end{figure}

Interfacial properties of a material are frequently extracted by fitting various models to the \textit{proxigram} composition profile. For example, sigmoid functions are well suited to model a symmetric interface ~\cite{ARDELL2012423,Ardell:2005fz} while manual thresholding more accurately extracts the properties of an asymmetric interface~\cite{Plotnikov:2014fs}. The accuracy of the interfacial properties impacts derivative material properties, such as interfacial free energy. Specifically, the interfacial free energy depends on the mean precipitate radius and the supersaturation of each element in the system, properties that APT datasets aptly capture. However, any interface identification or analysis requires determining the interface location through an iso-concentration analysis. Using the HED approach described here, we performed interface detection through compositional contrast rather than an arbitrarily defined concentration value. Therefore, the HED is agnostic with respect to models of an interface, to both detection and other properties, such as symmetry or asymmetry.

The resulting \textit{proxigram} from the layered structure generated by using MD is
shown in Figure~\ref{fig:MD_layered}, and those of the synthetic IVAS-generated data are shown in
Figure~\ref{fig:3DAPT_layered}. For the MD synthetic structure, since the Co concentration does not monotonically vary across the interface, adopting a spline fit and Co concentration thresholding will more accurately capture the interface properties as commonly practiced in the phase-field modeling community. With the concentration thresholding method, the interface thickness obtained by using the HED method is $2.3$~{\AA}, while that obtained from IVAS is approximately
$1.3$~{\AA}. We note that the interfacial width obtained for the same dataset using the
isoconcentration surface obtained from IVAS (Figure~\ref{fig:MD_layered}) yields an interface
thickness that is smaller than that obtained by using the HED approach. The far-field concentrations of Co on either side of the interface ($80.2\%$ and
$50.7\%$) are both accurately recovered by the \textit{proxigram} method, which provides
a quantitative validation of the interface obtained with our segmentation
approach. Far-field concentrations are calculated by averaging the composition after the interfacial contribution has diminished.

Properties of a symmetric interface may be modeled by using the following sigmoid function,
\begin{eqnarray}
\label{CurveEquation}
f(x)  = \frac{1}{2}(\rho_{1} + \rho_{2}) - \frac{1}{2}(\rho_{1} + \rho_{2}) * \tanh \left( \frac{x - a}{b} \right),
\end{eqnarray}
where $\rho_{1}$, $\rho_{2}$, $a$, and $b$ are fitting parameters that
correspond to the atomic densities of the two regions of the box, the interface position,
and the interfacial thickness, respectively. Direct measurement of the interfacial thickness of the MD dataset using Equation \ref{CurveEquation} on time-averaged histograms of the Co positions along the interfacial direction results in an interfacial thickness of $3.8$~{\AA}, whereas fitting Equation \ref{CurveEquation} to the HED \textit{proxigram} results in an interfacial thickness of $4.0$~{\AA}. This comparison suggests that the HED interface can closely capture the interface thickness, a capability that is important for calculating material properties that are particularly sensitive to measurement error, such as the coarsening rate constant.  

We also compare the \textit{proxigram} concentration profile obtained from the synthetic layered data
generated by using IVAS (Figure~\ref{fig:3DAPT_layered}). In this case, we use the interfacial
surface generated by using HED approach as well as that obtained by using IVAS.  
The \textit{proxigram} concentration profiles generated from IVAS use a bin size of 0.5~nm to average local composition fluctuations. As in the earlier case, we find that the concentrations on
either side of the interface are accurately obtained using both approaches.
However, even though the interfacial thickness is very close for
both approaches, the interfacial thickness obtaind by using the proposed HED approach is slightly larger than than that obtained by using IVAS, 
consistent with our observations for the synthetic MD-generated interface (Figure~\ref{fig:MD_layered}).

\section{Discussion}
\label{Sec:conclusions}

The ability to both rapidly and accurately identify interfaces, as well as to analyze
them qualitatively and quantitatively, is paramount in the analyses of APT
data, and also in general to tomographic investigations of multi-grain solid
structures \cite{Wither2014, CNUDDE20131, MOBUS200718, Midgley:2009kj}.  
Current approaches utilizing iso-concentration
surfaces in interface identification require subjective input and may not be
automated and scaled to high-performance computing to extract interfacial
properties.

We have proposed and demonstrated a digital image segmentation approach to segment the precipitate
and the matrix phases in superalloys, and we used that approach to obtain the interfaces. The segmentation
is accomplished by using supervised edge detection to accommodate the irregularity in the
morphologies of the two phases. Specifically, we utilize the hierarchically nested
edge detection approach that consists of fully convolutional networks
along with deep supervision. Furthermore, because the process of collecting and
manually labeling the interface in the APT data for training
the edge detection model can be cumbersome, labor-intensive, expensive, and time-consuming, we propose a
transfer learning approach to ameliorate this approach. Our approach utilizes the edge
detection features learned on the natural images, which have abundant label data,
and transfers that knowledge to segmentation in the atom-probe tomography
data. 

We demonstrated that our approach is qualitatively and quantitatively accurate, 
by comparing  the results of our approach with that of proprietary IVAS
software from Cameca Instrument Inc. for synthetic and experimental APT data of two-phase systems.
Our 
approach successfully segmented the two phases and identified the
interfaces with different geometrical features. The identified interfaces correspond well
with the qualitative visualization using 2D slices. In cases where the images are noisy and the interface is
not clear, we obtain \textit{thick} edges, which could be alleviated by using
approaches such as crisp edge Detection~\cite{wang2017deep}.

The approach proposed here demonstrates the power of machine learning techniques in the
analyses of APT data. It should also be readily applicable to analysis of
other tomographic data of multi-phase or multi-grain systems, such as X-ray
tomography of multi-phase systems or transmission electron microscopy tomography. By
using transfer learning, the fully convolutional network can be trained in
advance of experiments and be applied in real time. This may be especially valuable in situations where rapid data analysis during the experiment may provide valuable real-time
feedback to the experiment.

\section{Methods}
\subsection{Proximity Histogram Calculation}
\label{S:4}

The 3D point cloud data from the APT results consist of the
atomic positions and elemental identities; when these data are combined with the edge-detected
surface, a one-dimensional plot is obtained that represents the change in the
relative concentration of the chemical species as a function of the distance from
the surface. This plot is referred to as the proximity histogram, or \textit{proxigram}, and is acquired following the procedure outlined in Hellman et al.~\cite{Hellman:2000ca}
except that the iso-concentration surface is herein replaced by the interfacial
surface obtained by edge detection. 
Algorithm~\ref{proxigramcalc} outlines the approach used to obtain the {proxigrams}.

\begin{algorithm}[H]
\begin{algorithmic}[1]
\State Obtain the edge detected surface.
\State Triangulate the surface to obtain mesh and corresponding elemental surface normals.
\For {$i = 1$ to number of elements}
	\State Calculate the distance of each atom with elemental identity $i$ from the surface\;
	\State $Hist_i \gets \text{Obtain histogram of the distances using a bin width } B$\;
\EndFor
\State $Hist_{all} \gets \text{Repeat the histogram calculation using atoms from all elements}$.
\For {i = 1 to number of elements}
	\State $\textit{Proxigram}_i = Hist_i/Hist_{all}$\;
\EndFor	
\end{algorithmic}
\caption{{\it{Proxigram}} calculation}\label{proxigramcalc}
\end{algorithm}

\subsection{Data Acquisition and Preparation}
\label{Sec:Data_prep}

\subsubsection{Synthetic Dataset for Validation and Verification}
\label{Sec:Data_prep_MD}

To create a verifiable test case, we created a synthetic sample with a known
concentration using an (MD) approach---an atomistic simulation method, where the motion of atoms is
modeled using Newton's laws of motion. In these simulations, the forces
between atoms are modeled using empirical forcefield equations, and motion is
created by using various time integration schemes. The synthetic sample was first
created by inserting two different amorphous mixtures of Al and Co
into a box at opposite ends. The two phases were bridged with an amorphous
interface whose mixture was initially a linear gradient of 5\r{A}. Once the initial
structure was generated, it was heated under molecular dynamics using the
LAMMPS software package~\cite{LAMMPS}. This heating was done at a constant temperature of 2000~K for
100 picoseconds of simulation time in order to smooth out the interface. The
Embedded Atom potential of Pun et al. was used for the inter-atomic
forces~\cite{Pun2015AlCoEAM}. The MD synthetic structure contains one layer with an
80/20 Co to Al mixture and a second layer with a 50/50 mixture. The
total number of atoms from all the elements, Co and Al, was
16,000, and the dimensions of the volume enclosing them were $1000\times100\times100$~
\AA$^3$.

Similarly, a layered synthetic structure with the composition of a Co-based
superalloy (see below) was generated by using Cameca's IVAS analysis code. In order to
more closely emulate the structure obtained from APT, where atomic positions
deviate from the perfect crystalline lattice due to uncertainties from the
field-induced evaporation process \cite{Larson:2013vf}, the IVAS synthetic
dataset was artificially introduced with a random spatial deviation from the
theoretical atomic sites.

\subsubsection{APT Datasets}
\label{Sec:Data_prep_3DAPT}

Cobalt and aluminum superalloys were used as model materials for the experimental
APT datasets. The Co-based superalloy used is a ternary alloy with
8.8 at.\% Al and 7.3 at.\% of W. 
Coherent precipitates of the $\gamma'$-phase (L1$_2$) are formed in the
$\gamma$-phase (fcc) matrix with concentration differences in the two
phases following the bulk thermodynamic potentials. These concentration differences lead to concentration changes across
interfaces that can be used to identify them. The total number of atoms from all the
elements (Co, Al, and W) is 19,104,918, with the
dimensions of the enclosing volume being
$112\times90.5\times90.5$~nm$^3$. The  \texttt{concentration space} is
obtained for the Co atoms with a voxel size chosen as $0.5\times0.5\times
0.5$~nm$^3$. 

The Al superalloy studied herein is an Al-Er-Sc-Zr-Si alloy that is strengthened by ordered (L1$_2$) coherent
Al$_3$(Er,Sc,Zr) precipitates and has a concentration of 0.005 at\% Er, 0.02 at\% Sc, 0.07 at\% Zr, and 0.06 at\% Si. The total number of atoms from all the elements (Al, Er, Sc, Zr and Si) is 668,388 with the dimensions of the enclosing volume being 29.5 $\times$ 23.5 $\times$ 22.5 nm$^3$. The \texttt{concentration space}
is obtained for the Al atoms with a voxel size chosen as 0.5 $\times$ 0.5 $\times$ 0.5
nm$^3$.  

Atom-probe tomograms of the alloys were obtained by preparing nanotips and were analyzed
by using a Cameca's local-electrode atom-probe (LEAP) 4000X-Si
equipped with picosecond ultraviloet laser. During the pulsed laser illumination, surface atoms
are evaporated toward a two-dimensional position sensitive detector, thus constructing three-dimensional atomic tomograms of the specimens. Along with the time-of-flight measurements, the mass-to-charge ratio of the
atoms can be determined, providing chemical information for each individual atoms.
The experimental procedures and tomogram reconstruction conditions are detailed
in Refs.~\cite{Bocchini:2018ih,Erdeniz:2017ia}. The APT raw datasets were
processed by using IVAS for mass spectra analyses and
spatial reconstructions. After raw data processing, a position file that
contains the reconstructed species' spatial positions and mass-to-charge state
(m/n) ratios was obtained. Additionally, a range file that matches an m/n ratio to a particular element enable identification of
each atomic species~\cite{Larson:2013vf}.

\section*{Acknowledgments}
S.M., T.L. and S.S.K.R.S. were supported by the U.S. 
Department of Energy, Office of Science, under Contract No.
DE-AC02-06CH11357. The work by D.-W.C., D.N.S., and O.H. was performed under financial assistance award
70NANB14H012 from the U.S. Department of Commerce, National Institute
of Standards and Technology as part of the Center for Hierarchical
Material Design (CHiMaD). P.B. acknowledges support from the U.S. Department of Energy, Office of Science, Advanced Scientific Computing Research Early Career Research Program. We gratefully acknowledge the computing
resources provided on Bebop and Blues, high-performance computing clusters operated by the Laboratory Computing Resource Center at
Argonne National Laboratory.

\section*{Author Contributions}
\noindent S.M. formulated the proposed approach, developed and trained the machine learning models and extracted the proxigrams. 
The APT experiments and IVAS data processing were performed by D.-W.C. with guidance from D.N.S. The interatomic potentials for MD were developed by T.L. and S.S.K.R.S, T.L. 
performed the MD simulations with guidance by S.S.K.R.S.
P.B. provided guidance on development and training of the machine learning approach. 
O.H. developed the concept for the research, provided guidance on formulating the proposed approach, supervised and coordinated the research. 
S.M and D.-W.C. wrote the manuscript with input from all the authors.\\

\section*{Additional Information}
\noindent The authors declare no competing interests.

\bibliographystyle{apsrev4-1}
\bibliography{sample}

\end{document}